\documentclass[aps,prb,twocolumn,superscriptaddress]{revtex4}
\usepackage{epsfig}
\usepackage{graphicx}
\usepackage{amsmath}
\usepackage{amsfonts}
\usepackage{amssymb}
\usepackage{bm}
\usepackage{bbold}
\usepackage{epsfig}
\usepackage{graphicx}
\usepackage{color}
\usepackage{pictex}
\usepackage[english]{babel}
\begin{document}
\title{Local and nonlocal order parameters in the Kitaev chain}
\author{Gennady Y. Chitov}
\affiliation{Department of Physics, Laurentian University, Sudbury, ON,
P3E 2C6 Canada}
\affiliation{Department of Physics,
McGill University, Montr\'{e}al, QC, H3A 2T8 Canada }
\date{\today}

%
%
\begin{abstract}
We have calculated order parameters for the phases of the Kitaev chain with interaction and dimerization
at a special symmetric point applying the Jordan-Wigner and other duality transformations.
We use string order parameters (SOPs) defined via the correlation functions of the Majorana
string operators. The SOPs are mapped onto the local order parameters of some dual Hamiltonians and
easily calculated.
We have shown that the phase diagram of the interacting dimerized chain
comprises the phases with the conventional local order as well as the phases with nonlocal SOPs.
From the results for the critical indices we infer the 2D Ising universality class of criticality at the particular
symmetry point where the model is exactly solvable.
\end{abstract}
\maketitle

\textit{$\bullet$ Introduction and Motivation.--} In the Landau theory phases
are distinguished by different types of long-ranged order, or its absence. The order is described
by an appropriately chosen order parameter. In the original version of the theory the latter quantity is understood
as local. Also, a continuous phase transition is related to spontaneous breaking of system's symmetry
expressed via local parameters of the Hamiltonian. \cite{Landau5}

It might appear that various low-dimensional fermionic or spin systems as quantum spin liquids,
frustrated magnetics, topological and Mott insulators, etc, \cite{FradkinBook13,TI,Ryu10,Montorsi12} which lack conventional local long-ranged
order even at zero temperature, cannot be dealt with in the Landau framework. The new paradigm of topological order
(for a recent review and references \cite{Wen16}) seems to be taking over. In our recent work we made a strong claim
that the Landau formalism, although extended, remains instrumental even for nonconventional quantum orders. \cite{GT2017}
The formalism needs to be extended to incorporate nonlocal (string) operators, \cite{Kogut79,denNijs89}
string correlation functions, and string order parameters (SOPs). The appearance of nonlocal SOP is accompanied
by a hidden symmetry breaking. \cite{HiddenSSB} The local and nonlocal order frameworks are related by duality, and
become a matter of convenient choice of variables of the Hamiltonian. \cite{Kogut79,GT2017,ChenHu07,Xiang07,Nussinov}
In a sense this is analogous to description of a crystal using direct or reciprocal Bravais lattices.

Some additional aspects of quantum ordering quantified by, e.g., topological numbers, Berry phases,
entanglement \cite{FradkinBook13} are nor reducible to the parameters of the Landau theory. These quantities provide
rather complementary description and do not seem to be indispensable, since the information one can get from the spectrum,
correlation functions, and the order parameter suffice to determine the phase diagram and the universality classes of
the transitions it contains.

The above apologia of the Landau paradigm might be not very appealing, since its almost \textit{``unbelievable simplicity"}
is at odds with the fashion trend for \textit{``more complex things which are .... easier."} \cite{Pasternak}
The main goal of the present work is to explain the recently found phase diagram of the dimerized
interacting Kitaev model using ``simple" basics of the Landau framework.  The key elements of dealing with local
and nonlocal orders were worked out in \cite{GT2017} using mainly results for the Heisenberg spin ladders.\cite{UsLadd}
Now we present a straightforward application of the methods developed in \cite{GT2017}
for the Kitaev fermionic chain.

\begin{figure}[h]
\centering{\includegraphics[width=8.0 cm]{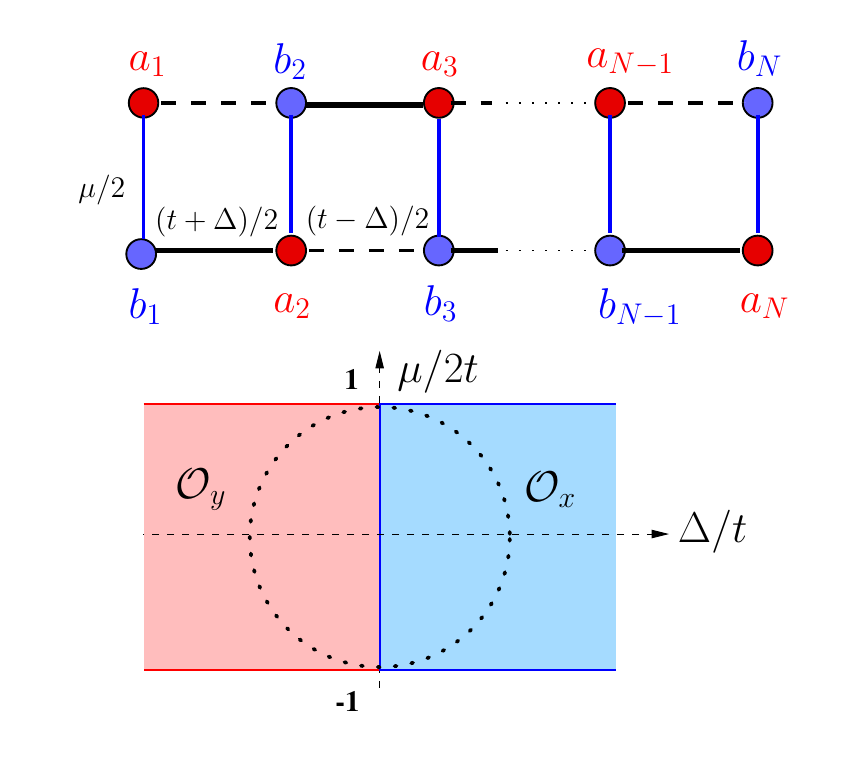}}
 \caption{(Color online) The Kitaev chain visualized as a two-leg ladder (upper panel) and its phase diagram (lower panel).
 The couplings in the ladder are shown according to the Hamiltonian \eqref{HKitMaj}. Two phases with nonzero SOPs are shown on
 the phase diagram along with the disorder line $(\mu/2t)^2+(\Delta/t)^2=1$ (dotted line).}
 \label{ChainLad}
\end{figure}

\textit{$\bullet$ Noninteracting Kitaev chain.--}
The Kitaev chain model of topological superconductor comprised of spinless fermions
is defined as\cite{Kitaev01}
\begin{eqnarray}
\label{HKit}
   H &=& \sum_{n=1}^{N} \Big\{ - \mu  \big( c^{\dag}_n c_n - \frac12 \big)
   -t \big( c^{\dag}_{n+1} c_n +   c^{\dag}_{n} c_{n+1} \big) \nonumber \\
 &+& \Delta  \big( c^{\dag}_{n+1}  c^{\dag}_{n} + c_n c_{n+1} \big) \Big\}~,
\end{eqnarray}
where $\mu$ is the chemical potential, $t$ is the hoping amplitude, and $\Delta$ is the (real)
superconducting gap. In terms of the Majorana operators
\begin{equation}
\label{Maj}
   a_n +i b_n  \equiv 2 c^{\dag}_n~.
\end{equation}
with the standard anticommutation relations
\begin{eqnarray}
\label{Anti}
  \{a_n,a_m\} &=& 2 \delta_{nm}, ~~ \{b_n,b_m\} = 2 \delta_{nm}, \nonumber \\
   \{a_n,b_m\} &=& 0
\end{eqnarray}
the Hamiltonian \eqref{HKit} reads
\begin{equation}
\label{HKitMaj}
  H = \frac{i}{2} \sum_{n=1}^{N} \Big\{ \mu a_n b_n- (t+\Delta) b_n a_{n+1} +(t-\Delta) a_n b_{n+1}
  \Big\}~.
\end{equation}
The Jordan-Wigner transformation \cite{Lieb61,Franchini2017} in the Majorana representation
\begin{equation}
\label{JWTMaj}
  \left(
    \begin{array}{c}
      \sigma_n^x \\
      \sigma_n^y \\
    \end{array}
  \right)
  =
  \left(
    \begin{array}{c}
      a_n \\
      b_n  \\
    \end{array}
  \right)
  \prod_{l=1}^{n-1} \big[ i a_l b_l \big]
\end{equation}
resulting in
\begin{eqnarray}
 \label{SigMaj}
  \sigma_{n}^{x}  \sigma_{n+1}^{x} &=& i b_n a_{n+1}~, ~~ \sigma_{n}^{y} \sigma_{n+1}^{y} = -i a_n b_{n+1}~, \nonumber \\
  \sigma_{n}^{z} &=& -i a_n b_n~,
\end{eqnarray}
maps the Kitaev model \eqref{HKitMaj} onto the $XY$ chain in the transverse field: \cite{Kitaev01,DeGottardi11}
\begin{equation}
\label{XYh}
   H =- \sum_{i=1}^{N} \bigg\{ \frac{t}{2}  \bigg[ \bigg(1+\frac{\Delta}{t} \bigg)
 \sigma_{i}^{x}\sigma_{i+1}^{x}+ \bigg( 1-\frac{\Delta}{t} \bigg) \sigma_{i}^{y}\sigma_{i+1}^{y} \bigg]
 + \frac12 \mu \sigma_{i}^{z} \bigg\}~,
\end{equation}
where $\sigma$-s are the Pauli matrices. The spectrum of this model
\begin{equation}
\label{EpsXYh}
 \varepsilon(k)= \pm 2t
 \sqrt{\Big( \frac{\mu }{2 t} -\cos k \Big)^2+ \Big(\frac{\Delta}{t} \Big)^2 \sin^2 k}~,
\end{equation}
its phase diagram and other properties are well known. \cite{Lieb61,Barouch71}
The properties of two models  \eqref{HKit} and \eqref{XYh} are identified from the correspondence
$J \leftrightarrow 2t$,  $\gamma  \leftrightarrow  \Delta/t$, and $h \leftrightarrow  \mu$, cf.
the definitions in \cite{Franchini2017}.  At ``strong field" $|h/J|=|\mu/2t|>1$ the Kitaev model does not have a nontrivial
order. To identify the order parameter at $|\mu/2t|<1$ we define the Majorana string operator:
\begin{equation}
\label{Ox}
  O_x(m) = \prod_{l=1}^{m-1} \big[ i b_l a_{l+1} \big]~.
\end{equation}
(By definition $O_x(1)=1$.) The SOP $\mathcal{O}_x$ is introduced as
\begin{equation}
\label{SOP}
 \mathcal{O}^2_{x} =\lim_{(n-m) \to \infty} |\langle O_x(m) O_x(n)  \rangle |~.
\end{equation}
As follows from relations \eqref{SigMaj}, the nonlocal Majorana string correlation function maps onto the two-point
spin correlation function of the dual $XY$ chain  \eqref{EpsXYh} (cf, e.g., \cite{Franchini2017}):
\begin{equation}
\label{OxMx}
  \langle O_x(m) O_x(n)  \rangle =   \langle  \prod_{l=m}^{n-1} \big[ i b_l a_{l+1} \big] \rangle
  =\langle \sigma_n^x \sigma_m^x \rangle ~.
\end{equation}
Introducing the longitudinal magnetization of the $XY$ chain as
\begin{equation}
 \label{Mx}
  m_{x}^{2} = \lim_{(m-n) \to \infty} |\langle\sigma_{n}^{x}\sigma_{m}^{x}\rangle|
\end{equation}
and using the results of \cite{Barouch71} we find the Majorana SOP at $|\mu/2t|<1$:
\begin{equation}
\mathcal{O}_{x} =
\left\{
\begin{array}{lr}
\displaystyle \sqrt{\frac{2}{1+\Delta/t}} \bigg(\Big(\frac{\Delta}{t}\Big)^2 \Big[1-\Big(\frac{\mu}{2t}\Big)^2 \Big] \bigg)^{1/8}~;
&  ~\Delta/t > 0 \\[0.5cm]
\displaystyle 0~; & ~\Delta/t < 0
\end{array}
\right. \label{OxRes}
\end{equation}
To probe the region $\Delta/t <0$ we define another Majorana string:
\begin{equation}
\label{Oy}
  O_y(m) = \prod_{l=1}^{m-1} \big[ - i a_l b_{l+1} \big]~.
\end{equation}
Similarly we find the SOP at $|\mu/2t|<1$:
\begin{equation}
\mathcal{O}_{y} =
\left\{
\begin{array}{lr}
0~;  &  ~\Delta/t > 0 \\
\mathcal{O}_{x}(-\Delta/t)~; & ~\Delta/t < 0
\end{array}
\right. \label{OyRes}
\end{equation}
Note that the SOPs $\mathcal{O}_{\alpha}$ ($\alpha=x,y$) and their dual magnetizations are the bulk parameters and their values
\eqref{OxRes}, \eqref{OyRes} are not sensitive to the choice of the ends of the strings (cf. \eqref{SOP}) as far
as the thermodynamic limit is taken and $(n-m) \to \infty$. We adapt the idea of DeGottardi and co-workers \cite{DeGottardi11}
to visualize the Kitaev chain as a two-leg ladder where two Majorana fermions comprizing a single Dirac fermion
reside on the rungs of this ladder, see Fig.~\ref{ChainLad}. Two string  Majorana operators yielding $\mathcal{O}_{x}$
and $\mathcal{O}_{y}$ correspond to two distinct snake-like paths on the ladder, cf. definitions \eqref{Ox} and \eqref{Oy}.
The string of maximal length  $O_{x/y}(N) $ for a chain of $N$ sites
does not include a pair of Majorana operators at the ends ($(a_1,b_N)/(b_1,a_N)$, resp.). Thus nonvanishing SOPs $\mathcal{O}_{x/y}$
signal correlations of the fermions along the chain and existence of two unpaired edge Majorana fermions.
(For details, see the Appendix.)
This is the feature associated with a topological order and that is why the phase with  $\mathcal{O}_{x/y} \neq 0$ is called
topological superconductor.\cite{Kitaev01,DeGottardi11,TI} The string $O_{z}(m)$ made out of pairs of Majorana fermions
residing on the rungs of the ladder is not useful at this point, since
$\langle \sigma_{n}^{z} \rangle = -i \langle a_n b_n \rangle \neq 0$ at $\mu \neq 0$.

The disorder line $(\mu/2t)^2+(\Delta/t)^2=1$ shown in Fig.~\ref{ChainLad} corresponds to a transition
(without gap closure) when the asymptotic behavior of string correlation functions changes.
The exponentially decaying functions acquire additional oscillations inside the circle.\cite{Barouch71}
The relation of this transition to the analytical properties of the model's partition function,
and the closely related wave functions of the zero-energy edge states are analyzed in the Appendix.

\begin{figure}[h]
\centering{\includegraphics[width=9.0 cm]{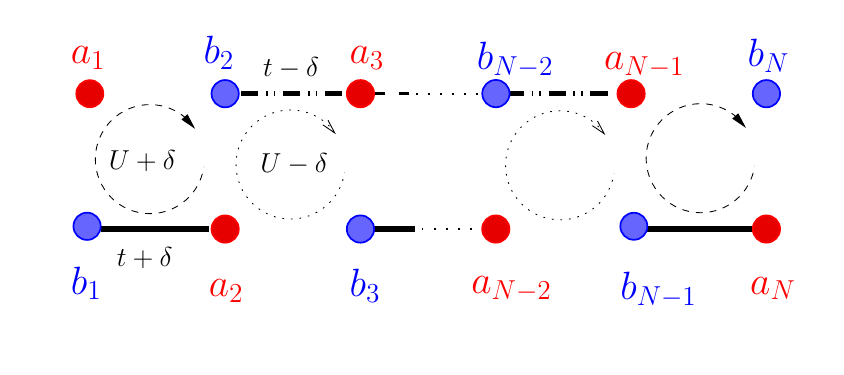}}
 \caption{(Color online) The interacting dimerized Kitaev chain visualized as a two-leg ladder.
The in-leg and plaquette couplings are shown according to the Hamiltonian \eqref{HxzM}.}
 \label{ChainLad2}
\end{figure}

\textit{$\bullet$ Dimerized interacting Kitaev chain.--} The Kitaev model was analysed also in the presence of
dimerization and interactions. \cite{EzawaNagaosa14,Zeng16,Miao17a,Ezawa17,Miao17b}
As shown by the recent exact solution of the interacting Kitaev chain at a special point,\cite{Miao17a} the interaction
brings about new phases. Interplay of interaction and dimerization makes the phase diagram of the model even richer.
Very recently the 1D dimerized interacting Kitaev models were proposed and solved virtually simultaneously at a special symmetric
point. \cite{Ezawa17,Miao17b}  Technically, the solution of the dimerized case is a straightforward extension of the earlier solution for the interacting model without dimerization. \cite{Miao17a} The models analyzed in \cite{Ezawa17,Miao17b} are slightly different (the version of Wang and co-workers \cite{Miao17b} does not have dimerization in interaction). We find the version of the model
proposed by Ezawa \cite{Ezawa17} slightly more convenient, and this is the Hamiltonian we will use in this paper:
\begin{eqnarray}
\label{HKitUdim}
   H = \sum_{n=1}^{N} &\Big\{& - \mu  \big( c^{\dag}_n c_n - \frac12 \big)
   -t_n \big( c^{\dag}_{n+1} c_n +   c^{\dag}_{n} c_{n+1} \big) \nonumber \\
 &+& \Delta_n  \big( c^{\dag}_{n+1}  c^{\dag}_{n} + c_n c_{n+1} \big) \nonumber \\
 &+& U_n \big( 2 c^{\dag}_{n+1} c_{n+1}-1 \big)   \big( 2 c^{\dag}_{n} c_{n}-1 \big) \Big\}~,
\end{eqnarray}
where
\begin{eqnarray}
\label{DimPar}
   t_n &=& t \big[1-(-1)^n \delta \big], ~~ \Delta_n= \Delta \big[1-(-1)^n \delta \big], \nonumber \\
   U_n &=& U \big[1-(-1)^n \delta \big]~.
\end{eqnarray}
Symmetries of the model \cite{Miao17a,Ezawa17} allow to assume $t>0$ and $\Delta >0$ without loss of generality,
and the dimerization parameter is bound $|\delta| \leq 1$.
The model \eqref{HKitUdim} is solved at the special point
\begin{equation}
\label{ExS}
  \Delta = t, ~ \mu =0.
\end{equation}
The Jordan-Wigner transformation maps the fermionic Hamiltonian  onto the spin model
\begin{eqnarray}
\label{HxzM}
 H &=& \sum_{n=1}^{N} \Big[-t_n i b_n a_{n+1} + U_n i a_n b_n i a_{n+1}  b_{n+1}\Big] \\
 \label{Hxz}
    &=& \sum_{n=1}^{N} \Big[-t_n  \sigma_{n}^{x}\sigma_{n+1}^{x} +
          U_n \sigma_{n}^{z}\sigma_{n+1}^{z}  \Big]~,
\end{eqnarray}
which after additional spin-rotational transformation becomes the well-known dimerized quantum $XY$
chain:\cite{Perk75,DelGamMod}
\begin{equation}
\label{Hxy}
H = \sum_{n=1}^{N} \Big[-t_n  \sigma_{n}^{x}\sigma_{n+1}^{x} +
          U_n \sigma_{n}^{y}\sigma_{n+1}^{y}  \Big]~.
\end{equation}
The Kitaev model in this particular exactly-solvable point is depicted as a two-leg ladder in
Fig.~\ref{ChainLad2}. Only the adjacent $b$ and $a$ operators are coupled along the legs, plus four Majorana
operators on each plaquette are coupled via alternating interaction.

At this point the critical properties of the model can be analyzed from fermionized Hamiltonian
\eqref{Hxy}. \cite{Ezawa17,Miao17b} Instead, to easily reveal the hidden order parameters
we will follow our recent analysis \cite{GT2017} and apply
the duality transformation:\cite{Fradkin78}
\begin{eqnarray}
  \sigma_{n}^{x} &=& \tau_{n-1}^{x}\tau_{n}^{x}
  \label{sigmaX} \\
  \sigma_{n}^{y} &=& \prod_{l=n}^{N} \tau_{l}^{z}~,
  \label{sigmaY}
\end{eqnarray}
where $\tau$-s obey the standard algebra of the Pauli operators, and they
reside on the sites of the dual lattice  which can be placed between the sites
of the original chain. This transformation maps the Hamiltonian \eqref{Hxy} onto a sum of two
decoupled 1D transverse-field Ising models \cite{Perk} defined on the even and odd
sites of the dual lattice:
\begin{eqnarray}
  H&=& H_e + H_o  \label{Hsum} \\
  H_e&=&  \sum_{l=1}^{N/2}
    -t(1+\delta) \tau_{2l-2}^{x} \tau_{2l}^{x}+ U(1-\delta)\tau_{2l}^{z}  \label{He} \\
  H_o&=&  \sum_{l=1}^{N/2}
    -t(1-\delta)\tau_{2l-1}^{x} \tau_{2l+1}^{x}+ U(1+\delta) \tau_{2l-1}^{z}  \label{Ho}
\end{eqnarray}
Such dual representation makes obvious the hidden $\mathbb{Z}_2 \otimes \mathbb{Z}_2$ symmetry of the
Kitaev chain.
The spectrum of the transverse Ising chain is well known, \cite{Lieb61,Barouch71,Pfeuty70} so
the eigenvalues of the Hamiltonian \eqref{Hsum} $\pm \varepsilon_{e/o}(k)$ read
\begin{equation}
\label{EpsQIM}
 \varepsilon_{e/o} (k)= \frac12 t(1 \pm \delta)
 \sqrt{\sin^2 k + \bigg( \cos k -\displaystyle \frac{U }{ t} \frac{1\mp \delta }{ 1 \pm \delta} \bigg)^2}~,
\end{equation}
in agreement with the earlier result of direct diagonalization. \cite{Ezawa17}
The lines of quantum criticality (gaplesness) for even and odd parts of the Hamiltonian \eqref{Hsum}
are:
\begin{equation}
\mathrm{even ~sector:}~~  \delta_{c,e} =
\left\{
\begin{array}{lr}
\displaystyle \frac{U/t-1}{U/t+1};  &  ~U/t > 0 \\[0.5cm]
 \displaystyle \frac{U/t+1}{U/t-1}; & ~U/t < 0
\end{array}
\right. \label{CrE}
\end{equation}
and in the odd sector:
\begin{equation}
 \delta_{c,o} =- \delta_{c,e}~.
\label{CrO}
\end{equation}
The chains are (ferromagnetically) ordered under the following conditions:
\begin{equation}
\label{TauxCond}
  \langle \tau^x_{e/o}\rangle \neq 0~~\mathrm{if}~~ \delta \gtrless \delta_{c,e/o}~.
\end{equation}

\begin{figure}[h]
\centering{\includegraphics[width=9.0 cm]{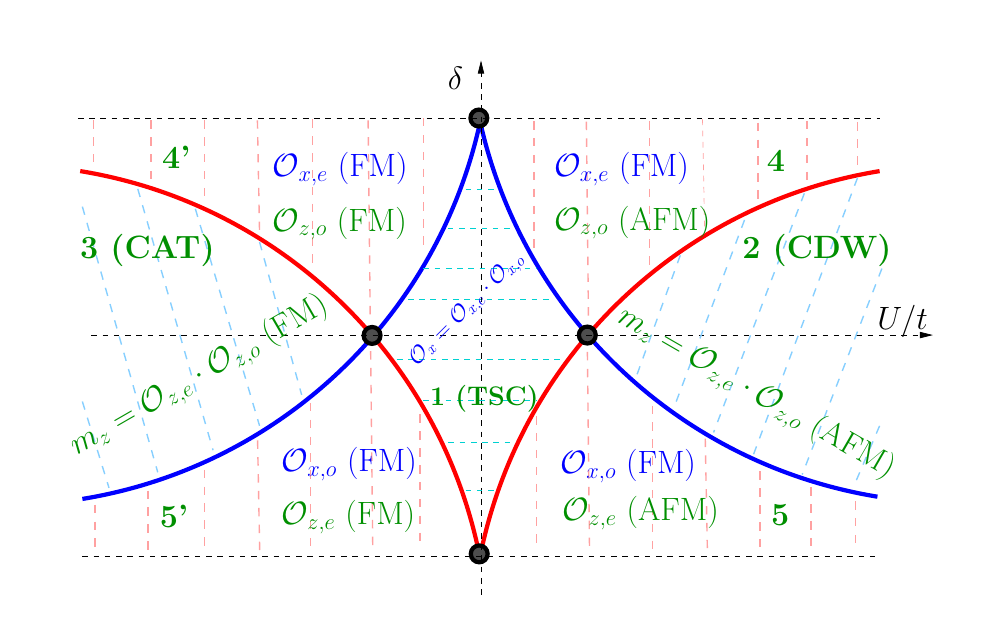}}
 \caption{(Color online) Phase diagram of the interacting dimerized Kitaev model with nonvanishing order parameter
 shown for each phase.}
 \label{PhDiag}
\end{figure}

The curves (\ref{CrE},\ref{CrO}) shown in Fig.~\ref{PhDiag} are the phase boundaries, and now we will establish the nature of the order
parameters characterizing each of the phases in the $(U/t, \delta)$ parametric plane. The phase diagram of the model was
found already in the earlier related work. \cite{Ezawa17,Miao17b} In particular, the winding numbers were calculated,
and it was shown that at least one number changes when a phase boundary is crossed, indicating thus topological phase transition(s)
along the lines (\ref{CrE},\ref{CrO}). This is not surprising, since the phase boundaries are the branching points of the spectra
\eqref{EpsQIM}. Our goal is to find the Landau-like order parameters for each phase of the diagram.
The phases 1-3 in Fig.~\ref{PhDiag} are continuous extensions of their counterparts found in \cite{Miao17a} at $\delta =0$.

\textit{Region 1 (Topological superconductor (TSC)):}
In this phase the $\mathbb{Z}_2$ symmetry is spontaneously broken in the even and odd sectors of the Hamiltonian \eqref{Hsum},
the ground state  is then 4-fold degenerate, and  $m_x =\langle \tau^x_e \rangle \langle \tau^x_o \rangle$. \cite{GT2017}
Then we easily find nontrivial Majorana SOP for this phase:
\begin{eqnarray}
\label{Otsc}
  \mathcal{O}_x &=& \displaystyle \Big[ \Big(1- \Big[\frac{U}{t}\frac{1-\delta}{1+\delta} \Big]^2\Big)
  \Big(1-\Big[\frac{U}{t}\frac{1+\delta}{1-\delta} \Big]^2 \Big) \Big]^{1/8}~, \nonumber \\[0.3cm]
   &~& \mathrm{at}~~ \{\delta_{c,e} <  \delta < \delta_{c,o} \} \cup \{ |U|/t<1 \}~.
\end{eqnarray}
This phase and its order parameter are smoothly connected to the corresponding phase of the free Kitaev chain
shown in Fig.~\ref{ChainLad}

\textit{Reqions 2 \& 3 (CDW \&  CAT):}
To calculate the order parameter(s) for those phases we apply the duality transformations
(\ref{sigmaX},\ref{sigmaY}) with the interchange $x \leftrightarrow y$. The Hamiltonian
\eqref{Hxy} maps again onto a sum of the even and odd transverse Ising chains as:
\begin{eqnarray}
  H_e&=&  \sum_{l=1}^{N/2}
   U(1-\delta) \tau_{2l-2}^{y} \tau_{2l}^{y}-t(1+\delta)\tau_{2l}^{z}  \label{HeY} \\
  H_o&=&  \sum_{l=1}^{N/2}
    U(1+\delta) \tau_{2l-1}^{y} \tau_{2l+1}^{y}-t(1-\delta) \tau_{2l-1}^{z}~, \label{HoY}
\end{eqnarray}
which have ferromagnetic or antiferromagnetic (depending on the sign of $U$) long-ranged order
\begin{equation}
\label{TauyCond}
  \langle \tau^y_{e/o}\rangle \neq 0~~\mathrm{if}~~ \delta \lessgtr  \delta_{c,e/o}~.
\end{equation}
The order parameter $m_z$ is defined via the density correlation function
\begin{eqnarray}
 4 \langle \delta n_i \delta n_j\rangle = \langle (2c^{\dag}_{i} c_{i}-1)(2c^{\dag}_{j} c_{j}-1)  \rangle \nonumber \\
 ~ \xrightarrow{(j-i) \to \infty}~[-\mathrm{sign}(U)]^{j-i} m_z^2~,
  \label{DD}
\end{eqnarray}
where the $z$-component is understood in terms of \eqref{HxzM} and \eqref{Hxz}
before the spin rotation. The nontrivial value of
\begin{equation}
\label{Mz}
  m_z= |\langle 2c^{\dag}_{n} c_{n}-1 \rangle |=|\langle a_{n} b_{n} \rangle |=
  |\langle \tau^y_e \rangle \langle \tau^y_o \rangle|~.
\end{equation}
is given by
\begin{eqnarray}
\label{MzEq}
  m_z &=& \displaystyle \Big[ \Big(1- \Big[\frac{t}{U}\frac{1+\delta}{1-\delta} \Big]^2\Big)
  \Big(1-\Big[\frac{t}{U}\frac{1-\delta}{1+\delta} \Big]^2 \Big) \Big]^{1/8}~, \nonumber \\[0.3cm]
   &~& \mathrm{at}~~ \{ \delta_{c,o} <  \delta < \delta_{c,e} \} \cup \{ |U|/t>1 \}~.
\end{eqnarray}
The phase with alternating density at $U/t>0$ is associated with the charge-density wave (CDW),
while the superposition of two differently homogeneously filled states (in our dual representation they are
dual even and odd sublattices with distinct ferromagnetic orders  $\langle \tau^y_e \rangle$
and  $\langle \tau^y_o \rangle$) at $U/t<0$ is called the CAT phase. \cite{Miao17a}
(Named after Schr\"{o}dinger's cat superposition state.)
Similarly to the TSC phase, the CDW and CAT phases have 4-fold degenerate ground states and
correspond to the completely broken $\mathbb{Z}_2 \otimes \mathbb{Z}_2$ symmetry.

\textit{Regions 4 \& 5:}
Now we introduce two types of rarefied strings \cite{IsingCasc} and define the even and odd Majorana string operators:
\begin{eqnarray}
  \label{Oxe}
  O_{x,e}(m) = \prod_{l=1}^{m} \big[ i b_{2l-1} a_{2l} \big]~, \\
  \label{Oxo}
  O_{x,o}(m) = \prod_{l=1}^{m} \big[ i b_{2l} a_{2l+1} \big]~.
\end{eqnarray}
The corresponding Majorana SOPs are defined similarly to \eqref{SOP}.
Using the Jordan-Wigner \eqref{SigMaj} and duality transformation (\ref{sigmaX})
we obtain the important relations
\begin{equation}
\mathcal{O}_{x,e/o} ^2 = \lim_{(R-L) \rightarrow\infty} \left<\tau_{L}^{x}\tau_{R}^{x}\right>~,
\label{SOPxeo}
\end{equation}
where the ends of the strings are chosen such that:
\begin{eqnarray}
  L=2n,~R=2m ~ &\longmapsto&~ \mathcal{O}_{\alpha,e}~,\nonumber \\
   L=2n-1,~R=2m-1 ~ &\longmapsto&~  \mathcal{O}_{\alpha,o}~.
  \label{LMpm}
\end{eqnarray}
This leads us easily to the nontrivial SOPs for the phases 4 and 5:
\begin{eqnarray}
\label{OxeoRes}
  \mathcal{O}_{x,e/o} &=& \displaystyle  \Big(1- \Big[\frac{U}{t}\frac{1\mp \delta}{1 \pm \delta} \Big]^2\Big)^{1/8}~, \nonumber \\[0.3cm]
   &~& \mathrm{at}~~ \{ \delta \gtrless \delta_{c,o} \} \cup \{ \delta \gtrless  \delta_{c,e} \}~.
\end{eqnarray}
As one can infer from definition \eqref{Oxe} and Fig.~\ref{ChainLad2}, the appearance of SOP $\mathcal{O}_{x,e} \neq 0$  signals nonvanishing average value of the rarefied string made out of Majorana dimers residing
on the same leg of the ladder and coupled by a ``plus" bond $t+ \delta$. Similarly, the nonvanishing $\mathcal{O}_{x,o}$ probes the rarefied strings made out the ``minus"  dimers coupled by $t- \delta$. Note that the SOP in the TSC phase $\mathcal{O}_{x}= \mathcal{O}_{x,e} \mathcal{O}_{x,o}$ is a superposition of these two rarefied strings where their nonzero averages overlap.
The ground states of the rarefied dimer phases 4 and 5 are two-fold degenerate, and nonvanishing SOPs $\mathcal{O}_{x,e/o}$ signal breaking of
one of the  $\mathbb{Z}_2 \otimes \mathbb{Z}_2$ symmetries, either in the even or odd sector of the dual Hamiltonian.

Another couple of SOPs in these two phases can be deduced from the local magnetizations \eqref{TauyCond} on the even/odd dual sublattices
\eqref{HeY} and \eqref{HoY} (cf. also \eqref{LMpm}):
\begin{eqnarray}
 \langle \tau_L^y \tau_R^y \rangle = \langle  \prod_{l=L+1}^{R} \big[- i a_l b_l \big]  \rangle \nonumber \\
 ~ \xrightarrow{(R-L) \to \infty}~[-\mathrm{sign}(U)]^{\frac{R-L}{2}} \mathcal{O}_{z,e/o}^2~,
  \label{Ozpm}
\end{eqnarray}
Analytically, we find
\begin{eqnarray}
\label{OzeoRes}
  \mathcal{O}_{z,e/o} &=& \displaystyle  \Big(1- \Big[\frac{t}{U}\frac{1 \pm \delta}{1\mp \delta} \Big]^2\Big)^{1/8}~, \nonumber \\[0.3cm]
   &~& \mathrm{at}~~ \{ \delta  \lessgtr  \delta_{c,o} \} \cup \{\delta \lessgtr \delta_{c,e} \} ~,
\end{eqnarray}
These two SOPs combine into the local order parameter (average density)
$m_z=\mathcal{O}_{z,e} \mathcal{O}_{z,o}$ in the phases CDW and CAT, where these even and odd SOP coexist.

\textit{$\bullet$ Conclusions.--} Using spin-fermion and spin-spin duality transformations we have calculated
order parameters for the phases of the noninteracting Kitaev chain and for the chain with interaction and
dimerization at a special symmetric point. The main building blocks we used are various string operators made
out of Majorana fermions. The string order parameters (SOPs) are defined by the asymptotes of the corresponding
string correlation functions. Using duality we show that the SOPs are local order parameters of the dual Hamiltonians
and are easily calculated. On the phase diagram  \cite{Ezawa17,Miao17b} of the interacting dimerized model
we have found the nonlocal order detected by the rarefied strings built from selected sets of the Majorana operators
(Phases 4 \& 5). Such rarefied SOPs coexist in the TSC phase and their overlap results in the SOP
which continuously evolves from the corresponding SOP of the noninteracting chain. The phases CAT and CDW possess
conventional local order parameters known from the analysis of the interacting model without dimerization. \cite{Miao17a}
Using the duality we have easily obtained the results for those local parameters as products of corresponding
overlapping SOPs. Each symmetry broken in a given phase of the model is identified with the spontaneous symmetry
breaking of the dual Hamiltonian(s). We have also related the phase transitions of the model (including the disorder
line) to zeros of its partition function. Form the results for the gaps and order parameters
we infer the critical indices $\nu=1$ and $\beta = 1/8$ of the 2D Ising universality class. \cite{Note}
This is valid of course only for the particular symmetry point in the parameter space considered where
the interacting model is equivalent to free fermions.

We expect the proposed SOPs to be operational to explore
the model's phase diagram  away from this special point, \cite{EzawaNagaosa14,Zeng16,Miao17a,Ezawa17,Miao17b}
when the fermionic interactions need to be dealt with. This can be done along the lines of our earlier related work
on spin ladders (i.e. interacting fermions). \cite{GT2017,UsLadd} Another interesting extension is the noninteracting Kitaev
chain with long-ranged superconducting pairing, which has a quite nontrivial phase diagram and very interesting critical
properties.\cite{LRpairing} These are very promising  and relatively straightforward directions for advancement of the
present formalism, which we relegate for future work.
%
%
\begin{acknowledgments}
The author thanks the Centre for Physics of Materials at McGill University
for hospitality. I am grateful to P.N. Timonin for valuable comments and for bringing
important papers to my attention and to V. Oudovenko for his help with software.
Financial support from the Laurentian University Research
Fund (LURF) is gratefully acknowledged.
\end{acknowledgments}
%

\begin {appendix}
\section{Critical and disorder lines, Lee-Yang zeros, and Majorana edge states
in the $XY$ model}\label{App}
%
%
Since the seminal papers by Yang and Lee \cite{LeeYang52} we are able to
rigorously relate phase transitions in a model to zeros of its partition function.
Such zeros in the $1D$  $XY$ model and its integrable deformations
which keep the Hamiltonian equivalent to free fermions, were analyzed in \cite{Tong06}.
(See also \cite{Tong12} for a follow-up work). In units of $J$ the Hamiltonian of the model
is
\begin{equation}
\label{XYDef}
   H =- \frac12 \sum_{i=1}^{N} \bigg\{ \frac12 \big(1+\gamma \big)
 \sigma_{i}^{x}\sigma_{i+1}^{x}+ \frac12 \big( 1-\gamma \big) \sigma_{i}^{y}\sigma_{i+1}^{y}
 + h \sigma_{i}^{z} \bigg\}~.
\end{equation}
The partition function
\begin{equation}
\label{Zxy}
   \mathcal{Z}(h, \gamma, T; \{ k \})= \prod_{k \in [0,2\pi]} e^{\beta  \varepsilon(k)/2}
   \Big( e^{-\beta  \varepsilon(k)} +1   \Big)~.
\end{equation}
has its zeros determined from the following equation:\cite{Tong06}
\begin{equation}
\label{LYZdef}
 \varepsilon(k)=
 \sqrt{( h -\cos k )^2+ \gamma^2 \sin^2 k}= i (2n+1) \pi T~,
\end{equation}
(we set $k_B=1$) resulting in the solution with a complex magnetic field
\begin{equation}
\label{LYT}
 h=\cos k \pm i \sqrt{ \gamma^2 \sin^2 k+ (2n+1)^2\pi^2 T^2}~.
\end{equation}
At zero temperature the Lee-Yang zeros are also zeros of the spectrum, and they are located on the ellipse in the plane
$h=h' +i h'' \in \mathbb{C}$:
\begin{equation}
\label{Ell}
 (h')^2 + \bigg(\frac{h''}{\gamma} \bigg)^2=1~.
\end{equation}
The Lee-Yang zeros located on the real axis of the complex magnetic plane are the points of quantum criticality.
For arbitrary $\gamma \neq 0$ this gives us to lines of the quantum phase transitions $h = \pm 1$, while
for $\gamma =0$ the ellipse collapses into the critical line $h \in [-1,1]$. Thus predictions for the phase boundaries
of the Lee-Yang formalism reproduce the results known from analysis of the correlation functions,\cite{Barouch71}
as it must be.

A more interesting question is to understand the origin of the transition on the disorder line
\begin{equation}
\label{DL}
 h^2 + \gamma^2 = 1~,
\end{equation}
where there is no gap closure, and the transition is detected by the change of asymptotic behavior of
the spin correlation functions: the exponentially decaying functions
acquire additional oscillations inside the circle \eqref{DL}.\cite{Barouch71}
The ground state energy
 \begin{equation}
\label{E_0}
 E_0 = - \frac{N}{2 \pi } \int_0^\pi dk \varepsilon(k) ~,
\end{equation}
is found in terms of elliptical functions \cite{Taylor85} and is shown \cite{Mac16} to be smooth and even infinitely differentiable
function on the boundary \eqref{DL}.

It is convenient to write
the spectrum $\varepsilon(k)$ in terms of the complex variable $z = e^{ik}$ as \cite{Franchini2017}
\begin{equation}
\label{epsZ}
 \varepsilon^2(z)=\frac{(1+\gamma)^2}{4}
 (z-\lambda_+) (z-\lambda_-)(z^{-1} -\lambda_+)(z^{-1}-\lambda_-)~,
\end{equation}
where
\begin{equation}
\label{lampm}
\lambda_\pm = \frac{h \pm \sqrt{h^2+\gamma^2-1}}{1+\gamma}~.
\end{equation}
The quantum phase transitions in the $XY$ model we discussed above are signalled by the zeros  $\lambda_\pm$
of the partition function  $\mathcal{Z}(h, \gamma, T \to 0; \{ z \})$ lying on the unit circle $|z|=1$.
Analytical continuation $k \rightarrowtail k +i k_0$ extends the product in the partition function \eqref{Zxy}
over the complex loop of arbitrary radius $e^{-k_0}$, which, in particular, can pass through  the roots $\lambda_\pm^{\pm1}$.
Thus, the latter are zeros of the zero-temperature partition function analytically continued onto the complex states $|z| \neq 1$.
On the other hand, $\lambda_\pm$ control the asymptotes of the correlation functions calculated from the Toeplitz determinants
\cite{McCoyBook, Franchini2017}.
The transition on the disorder line \eqref{DL} resulting in the oscillations corresponds to the points where $\lambda_\pm$
acquire imaginary parts and $\lambda_+=\lambda_-^\ast$. This is in a close analogy to the transition on the disorder line
in the classical Ising chain which corresponds to the Lee-Yang zeros in the range of complex parameters.\cite{IsingCasc}

There is even a more close analogy between the transitions on the disorder lines in the classical  \cite{Stephenson70,IsingCasc}
and the quantum transverse $XY$ chains. To reveal it one needs to find the zero-energy localized state in the ordered phase $h<1$ of the
model Hamiltonian \eqref{XYDef}. This problem was originally solved by Karevski \cite{Karevski00} whose transfer-matrix approach
we will follow. (The solution was repeated in more recent literature \cite{DeGottardi11}.) The Bogoliubov-de Gennes equation for the
Jordan-Wigner fermions in the direct space (cf. equations \eqref{HKit}, \eqref{HKitMaj} and \eqref{XYh} ) can be written as
\begin{equation}
\label{BdG}
    \left(
\begin{array}{cc}
  0 & \hat A - \hat B  \\
  \hat A + \hat B  & 0 \\
\end{array}
\right)~
\left(
  \begin{array}{c}
    \Phi_q\\
    \Psi_q \\
  \end{array}
\right)= \varepsilon(q)
\left(
  \begin{array}{c}
    \Phi_q\\
    \Psi_q \\
  \end{array}
\right)~,
\end{equation}
where $\hat A$ and $\hat B$ are $N \times N$  symmetric and antisymmetric matrices, respectively:
\begin{eqnarray}
  A_{ij} &=& 2h \delta_{i,j} +\delta_{i,j+1}+ \delta_{i,j-1} \\
  B_{ij} &=& - \gamma \delta_{i,j+1}+ \gamma \delta_{i,j-1} ~, \label{AB}
\end{eqnarray}
$\Phi_q=(\phi_q(1)....\phi_q(N))^T$ and $\Psi_q=(\psi_q(1)....\psi_q(N))^T$ are the $N$-component spinors
defining the Bogoliubov transformation of the Majorana fermions $(a_n,b_n)\mapsto (\alpha_q ,\beta_q)$ as
\begin{equation}
\label{Bogol}
2 \eta_q^\dag= \alpha_q +i \beta_q  =\sum_{n=1}^N \Big[\phi_q(n)a_n + i \psi_q(n) b_n \Big]~.
\end{equation}
In terms of the Bogoliubov fermions $\eta_q$ the Hamiltonian is diagonal:
\begin{equation}
\label{Heta}
  H=\sum_k \varepsilon(k) \big[\eta_k^\dag \eta_k - \frac12 \big] ~.
\end{equation}
The energy of the first excited singe-particle state $|1\rangle=\eta_1^\dag |GS\rangle$ vanishes in the thermodynamic limit and
this state becomes degenerate with the ground state in the ordered phase $h<1$. The wave function $\Phi_1$ of the Majorana
fermion in this state is found via iteration of the ``transfer matrix"
\begin{equation}
\label{T}
\hat T =\left(
          \begin{array}{cc}
            \frac{2h}{1+\gamma } & \frac{1-\gamma}{1+\gamma} \\
            -1 & 0 \\
          \end{array}
        \right)
\end{equation}
as
\begin{equation}
\label{Phin}
  \phi_1(n+1) = (-1)^n (\hat T^n)_{11} \phi_1(1) ~.
\end{equation}
The roots \eqref{lampm} also happen to be the eigenvalues of the ``transfer matrix". The latter can be written
via two orthogonal idempotent operators (projectors) $\hat{\mathcal{P}}_\pm$ as \cite{Lancaster69}
\begin{equation}
\label{Tpm}
  \hat{T} =\lambda_+  \hat{\mathcal{P}}_+  + \lambda_- \hat{\mathcal{P}}_- ~,
\end{equation}
where
\begin{equation}
\label{Epm}
  \hat{\mathcal{P}}_\pm \equiv \pm \frac{\hat T -\lambda_\mp \hat{\mathbb{1}} }{\lambda_+ - \lambda_-}~.
\end{equation}
Then  $\hat{T}^n =\lambda_+^n  \hat{\mathcal{P}}_+  + \lambda_-^n \hat{\mathcal{P}}_- $, and
we recover the result of Karevski: \cite{Karevski00}
\begin{equation}
\label{PhinExpl}
  \phi_1(n+1) = (-1)^n    \frac{\lambda_+^n -\lambda_-^n }{\lambda_+ -\lambda_-}\phi_1(1) ~.
\end{equation}
The wave function is delocalized when $h>1$, while in the ordered phase when
$h<1$ it is localized near the left edge of the chain $n=1$, and the probability density
for this zero-energy edge Majorana fermion exponentially decays with the distance
$\phi_1^2(n \gg 1) \propto \exp(-\kappa n)$. The inverse penetration depth $\kappa =2|\log \lambda_+|$
is also the inverse bulk correlation length of the spin correlation function. \cite{Barouch71}
Inside the disorder circle \eqref{DL} $\lambda_+=\lambda_-^\ast$ with $|\lambda_\pm|^2=(1-\gamma)/(1+\gamma)$,
the edge-state wavefunction \eqref{PhinExpl} acquires incommensurate oscillations on the top of the exponential decay.
The disorder line corresponds to a ``weak" continuous phase transition where the correlation length stays finite.
As in the classical Ising chain,\cite{Stephenson70} it demonstrates a cusp at the critical point. One can check that
$\kappa$ as a function of field, approaches its maximum on the disorder line with an infinite slope and stays
constant $\kappa=- \log \frac{1-\gamma}{1+\gamma}$ in the oscillating phase. It is natural to identify the
localized zero-energy edge-state with the Majorana fermion ($a_1$) ``missed" by the ordered Majorana string
($\mathcal{O}_x \neq 0$) discussed in the main text on the free Kitaev chain.

Since $\lambda_\pm (-\gamma) \lambda_\mp (\gamma) =1$, the second solution of \eqref{BdG} follows easily
\begin{equation}
\label{PsinExpl}
  \psi_1(n+1) = (-1)^n    \frac{\lambda_-^{-n} -\lambda_+^{-n} }{\lambda_-^{-1} -\lambda_+^{-1}}\psi_1(1) ~.
\end{equation}
In the ordered phase this wave function localized near the right edge \cite{DeGottardi11}
$\psi_1(n \gg 1) \sim \psi(N)) \exp(-(N-n) |\log \lambda_-|)$, acquires oscillations after crossing the disorder line.
It can be related to the zero-energy edge-state
of the Majorana fermion $b_N$. In the range of negative $\gamma$ the solutions interchange in the obvious way,
along with $(a_1,b_N) \leftrightarrow (b_1,a_N)$.

\end{appendix}

%

%
%
\end{document}